\documentclass[aps,prb,noeprint,preprint,a4paper,superscriptaddress]{revtex4-2}
	\usepackage{graphicx}
	\usepackage{dcolumn}
	\usepackage{bm}
	\usepackage{color}
	\usepackage{multirow}
	\usepackage[pdfborder={0 0 0},colorlinks=True,linkcolor=blue,urlcolor=blue,citecolor=blue]{hyperref}

	\newcommand{\cfo}{\(\text{CoFe}_2\text{O}_4\)} 
	\newcommand{\nfo}{\(\text{NiFe}_2\text{O}_4\)} 
	\newcommand{\alo}{\(\text{Al}_2\text{O}_3\)}  
	\newcommand{\feo}{\(\text{Fe}_3\text{O}_4\)} 
	\definecolor{dgreen}{rgb}{0,0.7,0}
	\definecolor{brown}{rgb}{0.87,0.46,0.1}
	\newcommand{\aka}[1]{{\textcolor{black}{#1}}}			
	\newcommand{\aoi}[1]{{\textcolor{black}{#1}}}			

	\definecolor{dgreen}{rgb}{0,0.7,0}
	
	\graphicspath{{./fig/}}

\begin{document}
	\title{Origin of magnetically dead layers in spinel ferrites \textit{M}Fe\textsubscript{2}O\textsubscript{4} grown on \(\text{Al}_2\text{O}_3\): Effects of post-deposition annealing studied by XMCD}

		\author{Yosuke Nonaka}
		\email[Corresponding author: ]{nonaka.hb6.yohsuke@jp.nipponsteel.com}
		\altaffiliation[Current affiliation: ]{Nippon Steel Corporation, Steel Research Laboratories.}
		\affiliation{Department of Physics, the University of Tokyo, Bunkyo-ku, Tokyo 113-0033, Japan}
		\author{Yuki K. Wakabayashi}
		\affiliation{Department of Electrical Engineering and Information Systems, The University of Tokyo, 7-3-1 Hongo, Bunkyo-ku, Tokyo 113-8656, Japan}
		\author{Goro Shibata}
		\affiliation{Department of Physics, the University of Tokyo, Bunkyo-ku, Tokyo 113-0033, Japan}
		\affiliation{Materials Sciences Research Center, Japan Atomic Energy Agency (JAEA), Sayo, Hyogo 679-5148, Japan}
		\author{Shoya Sakamoto}
		\affiliation{Department of Physics, the University of Tokyo, Bunkyo-ku, Tokyo 113-0033, Japan}
		\affiliation{The Institute for Solid State Physics, the University of Tokyo, Kashiwa, Chiba 277-8581, Japan}
		\author{Keisuke Ikeda}
		\affiliation{Department of Physics, the University of Tokyo, Bunkyo-ku, Tokyo 113-0033, Japan}
		\author{Zhendong Chi}
		\affiliation{Department of Physics, the University of Tokyo, Bunkyo-ku, Tokyo 113-0033, Japan}
		\author{Yuxuan Wan}
		\affiliation{Department of Physics, the University of Tokyo, Bunkyo-ku, Tokyo 113-0033, Japan}
		\author{Masahiro Suzuki}
		\affiliation{Department of Physics, the University of Tokyo, Bunkyo-ku, Tokyo 113-0033, Japan}
		\author{Arata Tanaka}
		\affiliation{Graduate School of Advanced Science of Matter, Hiroshima University, Higashi-hiroshima, Hiroshima 739-8530, Japan}
		\author{Masaaki Tanaka}
		\affiliation{Department of Electrical Engineering and Information Systems, The University of Tokyo, 7-3-1 Hongo, Bunkyo-ku, Tokyo 113-8656, Japan}
		\affiliation{Center for Spintronics Research Network, Graduate School of Engineering, The University of Tokyo, 7-3-1 Hongo, Bunkyo-ku, Tokyo 113-8656, Japan}
		\author{Atsushi Fujimori}
		\affiliation{Department of Physics, the University of Tokyo, Bunkyo-ku, Tokyo 113-0033, Japan}
		\affiliation{Center for Quantum Science and Technology and Department of Physics, National Tsing Hua University, Hsinchu 30013, Taiwan}
	%
	\date{\today}
	\begin{abstract}
		We study the electronic and magnetic states of as-grown and annealed \textit{M}Fe\textsubscript{2}O\textsubscript{4}(111)/\alo(111) (\textit{M}=Co, Ni) thin films with various thicknesses grown on Si(111) substrates with the \(\gamma\)-\alo(111) buffer layers by using x-ray absorption spectroscopy (XAS) and x-ray magnetic circular dichroism (XMCD), to investigate magnetically dead layers in these films.
		Although the magnetically dead layers in the as-grown samples are formed near the interface with the \alo\ buffer layer, we reveal that ferrimagnetic order is partially recovered by post-deposition annealing at 973 K for 48 hours in air.
		By analyzing the line shapes of the XAS and XMCD spectra, we conclude that\aoi{, in the dead layers,} there are a significant number of vacancies at the \(T_d\) sites of the spinel structure, which may be the microscopic origin of the degraded ferrimagnetic order in the \textit{M}Fe\textsubscript{2}O\textsubscript{4} thin films.
	\end{abstract}
	
	\maketitle
	
	\section{Introduction}
		Electronics have achieved great success by controlling the charge degree of freedom of electrons in semiconductors. 
		By adding the spin degree of freedom to the existing electronics, a new field of electronics \aoi{known as spintronics} is being developed.
		Spintronics devices have several advantages over existing electronic devices, e.g., \aoi{non-volatilitiy, low energy consumption, high-speed, infinite endurance} \cite{Wolf2001}.
		To establish next-generation electronics with semiconductor-based spintronics, injection of spin-polarized carriers into semiconductors should be realized at room temperature by utilizing a magnetic spin injector formed on the semiconductor.
		The spin filter is a promising technology to inject highly spin-polarized current into semiconductors.
		Insulating spinel-ferrite thin films are attracting high attention as a promising material for spin filters because they have the down-spin and up-spin conduction bands \aoi{at different energies} [\aoi{the} 3\textit{d-t}\textsubscript{2g} \aoi{band} of Fe\(^{3+}(O_h)\) and \aoi{the} 3\textit{d-e} \aoi{band} of Fe\(^{3+}(T_d)\), \aoi{as Fig. \ref{fig:structanneal}(a)}] \cite{Szotek2006}, and their N\'eel temperatures are much higher than room temperature (790 K for \cfo\ \cite{Suzuki1996,Ramos2007a,Sawatzky1968}, and 850 K for \nfo\ \cite{Szotek2006,Matzen2014}). 
		However, the experimentally obtained spin-filtering efficiency of ferrite-based tunnel barriers still remains considerably lower than the theoretical value of \(100\%\) \cite{Matzen2012,Moodera2007,Moussy2013,Luders2006}.
		\aoi{In} recent investigation, possible causes for this low spin-filtering efficiency \aoi{have been} proposed.
		One is that midgap impurity states are induced by structural and/or chemical defects \cite{Ramos2007a,Szotek2006,Takahashi2010}.
		For example, in \cfo, the presence of Co\(^{3+}\) degrades the spin filter due to the formation of up-spin states in the middle of the band gap. 
		Another possibility is that layers with degraded magnetic properties are formed in the interfacial region, which is often called magnetically dead layers.
		They are well known \aoi{for} epitaxial \feo\ thin films grown on MgO substrates, and are thought to originate from antiphase boundaries and their antiferromagnetic interdomain exchange interaction \cite{Margulies1996,Margulies1997,Voogt1998,Eerenstein2002,Moussy2004}.
		In addition, a recent study on epitaxial \feo\ thin films grown on MgO substrates revealed that the first \feo\ monolayer lacks Fe\(^{3+}\) ions at the tetrahedral sites \cite{Chang2016}, \aoi{resulting} in a loss of the ferrimagnetic order due to missing superexchange paths.

		In recent works \cite{Wakabayashi2017, Wakabayashi2018, Nonaka2021}, the magnetic properties of \(\textit{M}\text{Fe}_2\text{O}_4\)(111)/\alo(111)/Si(111) (\textit{M}=Co, Ni) structures \cite{Bachelet2014} were studied using the element-specific probes of x-ray absorption spectroscopy (XAS) and x-ray magnetic circular dichroism (XMCD).
		The authors revealed that the \aoi{decrease of} the inversion parameter \textit{y} \aoi{, defined by} \(= [\textit{M}_{1-y}\text{Fe}_y]_{T_d}[\text{Fe}_{2-y}\textit{M}_y]_{O_h}\text{O}_4\) is correlated with the formation of magnetically dead layers.
		The inversion parameter \textit{y}, which is equal to unity in the ideal inverse spinel, represents the regularity of the cation distribution in the spinel structure.
		It was found that the reduced thickness of the \cfo\ film leads to a significant reduction in both \(y\) and ferrimagnetic ordered moment.
		It was also found that \nfo\ has \aoi{a} larger \(y\) (\(=0.79 \text{--} 0.91\)) than \cfo\ (\(y=0.54\text{--}0.75\)), indicating that Ni has a higher \aoi{\(O_h\)-}site selectivity than Co.
		Considering that the cation distribution in \(\textit{M}\text{Fe}_2\text{O}_4\) can be modified by post-deposition annealing \cite{Sawatzky1968,Hu2000}, it is expected that annealing recovers the \aoi{ferri}magnetic \aoi{behavior} of the interfacial region degraded by the formation of the magnetically dead layers.
		In the present study, we investigate the electronic and magnetic states of as-grown and annealed \(\textit{M}\text{Fe}_2\text{O}_4\)(111)/\alo(111)/Si(111) (\textit{M}=Co, Ni) thin films using XAS and XMCD.
		\aoi{We} reveal that the degraded magnetic properties of the magnetically dead layer can be partially recovered by annealing through the redistribution of the cations. 
		Furthermore, we find that there are a significant number of vacant \(T_d\) sites in the magnetically dead layers, and that such vacancies may degrade the ferrimagnetic order in the \(\textit{M}\text{Fe}_2\text{O}_4\) thin films.
	\section{Experimental Methods}
\begin{figure}
			\centering
			\includegraphics[width=0.7\linewidth]{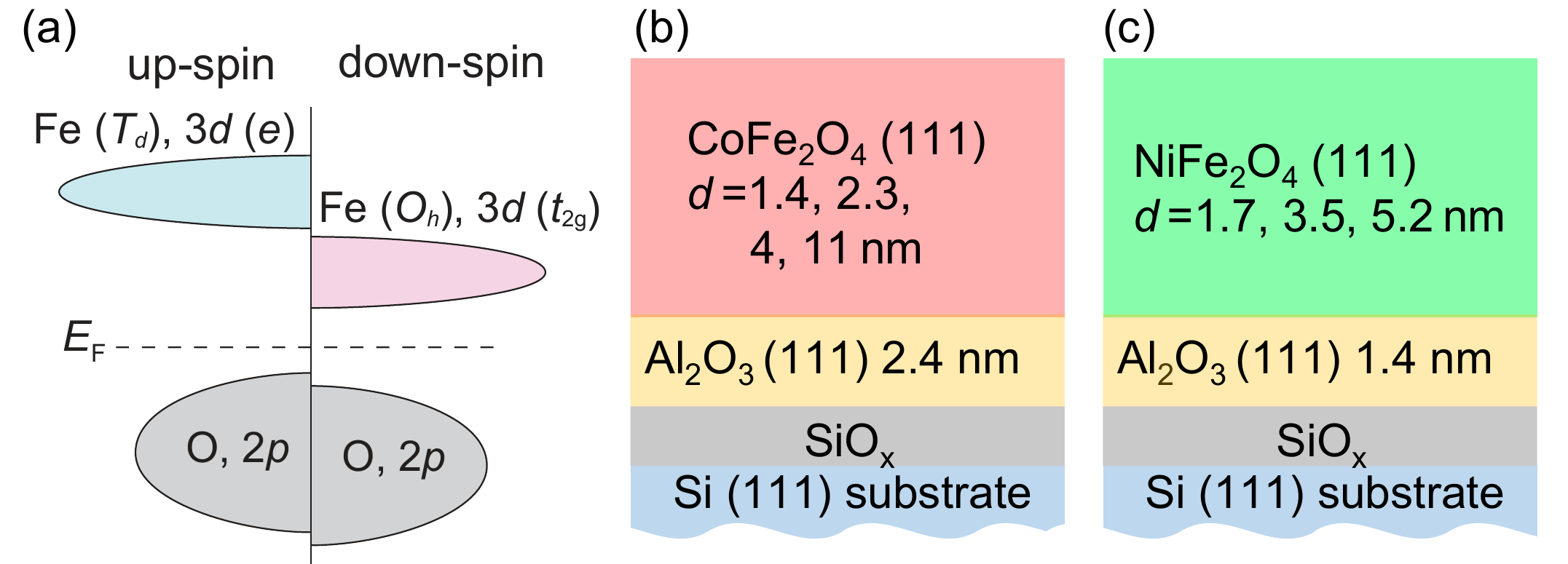}
			\caption{
				Schematic \aoi{description} of the \aka{electronic structure of \(\textit{M}\text{Fe}_2\text{O}_4\) (\textit{M} = Co or Ni) and the sample structures.} 
				(a) \aka{Density of states of the valence-band top and conduction-band bottom for \(\textit{M}\text{Fe}_2\text{O}_4\), reproduced from Ref. \onlinecite{Wakabayashi2018}.} 
				(b) Cross-sectional illustration of the \cfo(111)/\alo(111)/Si(111) structure, adapted from Ref. \onlinecite{Wakabayashi2017}.
				(c) Cross-sectional illustration of the \nfo(111)/\alo(111)/Si(111) structure, adapted from Ref. \onlinecite{Wakabayashi2018}.
			}\label{fig:structanneal}
		\end{figure}

		\cfo(111) and \nfo(111) thin films with various thicknesses were epitaxially grown on \alo(111) buffer layers formed on Si(111) substrates using the pulsed laser deposition (PLD) method.
		Schematic illustrations of the sample structures are shown in Figs. \ref{fig:structanneal}(b) and \ref{fig:structanneal}(c).
		2.4 nm- and 1.4 nm-thick \(\gamma\)-\alo(111) buffer layers were adopted for \cfo\ and \nfo\ thin films, respectively. 
		The thicknesses of the films were \textit{d} = 1.4, 2.3, 4, and 11 nm for \cfo, and \textit{d} = 1.7, 3.5, and 5.2 nm for \nfo.
		In order to avoid charging up of the samples during the XAS and XMCD measurements, we used highly phosphorus-doped Si(111) substrates with \aoi{a} low electrical resistivity 2 m\(\Omega\)cm.
		For the epitaxial growth of the \(\gamma\)-\alo\ buffer layers on the Si substrates, we used solid-phase reaction of Al and \(\text{SiO}_2\).
		More detailed description of sample preparation and characterization are provided in Refs. \onlinecite{Wakabayashi2017, Wakabayashi2018}.
		Some of the \cfo\ and \nfo\ thin films were annealed in order to redistribute the cations in the thin films, aiming at the recovery of the inversion parameter \(y\).
		In order to suppress the oxidation of the Si substrate, the post-deposition annealing temperature was set to 973 K because the oxidation rate of Si rapidly increases particularly above 1073 K \cite{Deal1965}. 
		The films were annealed at 973 K for 48 hours in air.
		Since higher inversion parameters of \cfo\ have been reported in slowly cooled bulk crystals than in rapidly quenched ones \cite{Sawatzky1968}, we set a slow cooling rate of 1 K/min down to 673 K and then the samples were naturally cooled down to room temperature.
		
		XMCD measurements were performed at beamline BL-16A of Photon Factory, High Energy Accelerator Research Organization (KEK-PF).
		A magnetic field of 5 T was applied parallel to the incident x rays.
		The measurements were done at room temperature (\(\sim300\) K).
		The absorption signals were detected in the total-electron-yield (TEY) mode.
		Since the probing depth in the TEY mode is a few nm \cite{Thole1985, Frazer2003}, the XAS and XMCD spectra of a few nm-thick films reflect the electronic and magnetic properties of the entire \(\textit{M}\text{Fe}_2\text{O}_4\) film including the magnetically dead layers near the spinel/\alo\ interface. 
		By increasing the film thickness to \(\gtrsim 5 \text{ nm}\), the spectra more strongly reflect the signals \aoi{away from the interface}.
		The XMCD signals were collected by switching the photon helicity at every energy point.

		In order to analyze the XAS and XMCD spectra quantitatively, we performed configuration-interaction (CI) cluster-model calculation \cite{Bocquet1992}.
		The model is a good approximation for localized electron systems like the insulating spinel ferrites.
		In the present calculation, we used \aoi{an} Xtls version 8.5 program \cite{Tanaka1994}.
		In choosing parameter values for the calculation, we adopted the following empirical rules : (1) The ratio between the on-site Coulomb energy between 3\textit{d} electrons \textit{U\textsubscript{dd}} and the attractive 2\textit{p} core hole-3\textit{d} electron Coulomb energy \textit{U\textsubscript{dc}} was fixed to \(U_{dc} / U_{dd} \sim 1.3 \) \cite{Chen2004, Hufner2003}.
		(2) The ratio between Slater-Koster parameters \((pd\sigma)\) and \((pd\pi)\) was fixed to \((pd\sigma) / (pd\pi) = -2.17\) \cite{Harrison1980}.
		The hybridization strength between O 2\textit{p} orbitals \(T_{pp}\) was fixed to 0.7 eV for the \(O_h\) site and 0 eV for the \(T_d\) site \cite{Tanaka1994, Chen2004}.
		The Slater integrals were set to be 80\% of the Hartree-Fock values. 
		Thus, the crystal-field splitting 10\textit{Dq}, the charge-transfer energy \(\Delta\), \(U_{dd}\), and \((pd\sigma)\) were treated as adjustable parameters. 
		As for the Fe cations, the parameters were adjusted to reproduce all the measured Fe \(L_{2,3}\)-edge spectra by the weighted sum of the calculated spectra for the Fe\(^{2+}(O_h)\), Fe\(^{3+}(T_d)\), and Fe\(^{3+}(O_h)\) ions.
		Here, parameters for Fe\(^{2+}(O_h)\) were adopted from the previous report on \nfo\ \cite{Wakabayashi2018}.
		The adjusted parameter values are listed in Table \ref{tab:para}. 
		\begin{table}
			\caption{Parameter values for the CI cluster-model calculations used in the present study in units of eV. Parameters for Fe\(^{2+}(O_h)\) were adopted from Ref. \onlinecite{Wakabayashi2018}.}
			\begin{center}
				\begin{tabular}{c c c c c} 
					\hline
					\hline
							&\(\Delta\)	&10\(Dq\)	&\((pd\sigma)\)	&\(U_{dd}\)	\\
					\hline
					Fe\(^{3+}(O_h)\)&3.0	&0.8	&1.5	&7.0	\\
					Fe\(^{3+}(T_d)\)&2.5	&-0.5	&1.8	&6.0	\\
					Fe\(^{2+}(O_h)\)&6.5	&0.9	&1.4	&6.0	\\
					\hline
					\hline
				\end{tabular}
			\label{tab:para}
			\end{center}
		\end{table}
		As in our previous works \cite{Wakabayashi2017,Wakabayashi2018}, we fitted the weighted sum of the calculated XAS and XMCD spectra for the Fe\(^{2+}(O_h)\), Fe\(^{3+}(T_d)\), and Fe\(^{3+}(O_h)\) ions to the experimental ones.
		From this analysis, we obtained the relative amount of the three Fe components and thereby the inversion parameter \textit{y}.
	\section{Results and Discussion}	
		\subsection{\cfo\ thin films}\label{aneal_cfo}
			\begin{figure}
				\centering
				\includegraphics[width=0.5\linewidth]{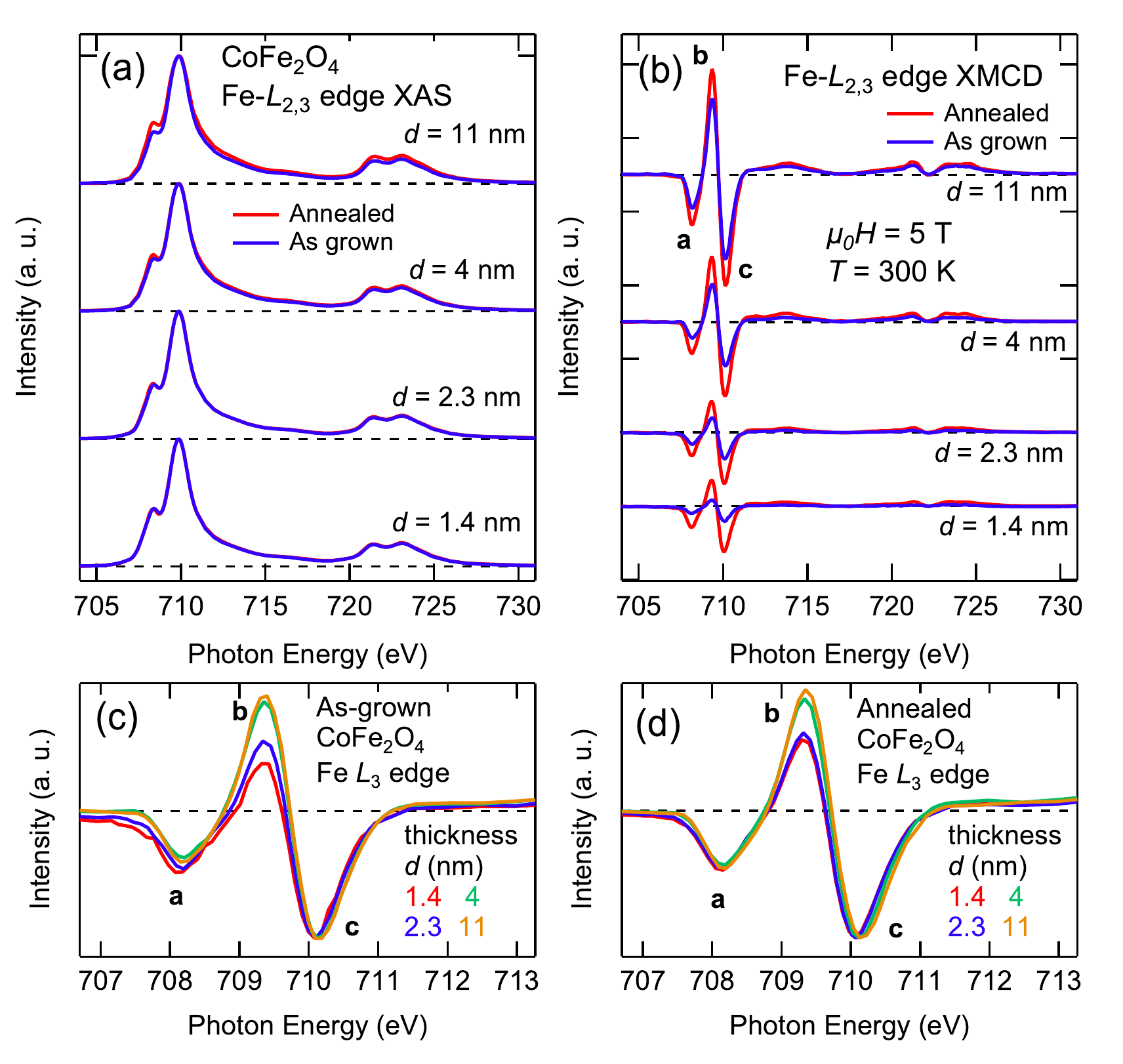}
				\caption{
					Fe \textit{L}\textsubscript{2,3}-edge XAS and XMCD spectra of the as-grown and annealed \cfo\ thin films with various thicknesses. 
					(a), (b) Fe \textit{L}\textsubscript{2,3}-edge XAS and XMCD spectra. In panel (a), a linear and a two-step backgrounds have been subtracted from the raw XAS spectra. 
					The two-step backgrounds are composed of two \aoi{arctangent} functions located at the peak positions of \(L_3\) and \(L_2\) white lines. The height of the \(L_3\) and \(L_2\) steps are set to \(\frac{2}{3}\) and \(\frac{1}{3}\) of the average height around 740 eV, respectively.
					(c), (d) Magnified views of the Fe \textit{L}\textsubscript{3}-edge XMCD spectra normalized to the height of peak \textbf{c}.
					The XMCD intensities decrease with \aoi{decreasing} thickness in the as-grown films [(b) and (c)], especially at peak \textbf{b}, reflecting the magnetically dead layers near the interface. Both the entire XMCD intensities and the relative intensity at peak \textbf{b} are partially recovered by post-deposition annealing particularly in thinner films [(b) and (d)].
				}\label{fig:cfo_fe}
			\end{figure}
			
			Figure \ref{fig:cfo_fe} shows the Fe \(L_{2,3}\)-edge XAS (\(\mu^++\mu^-\)) and XMCD (\(\mu^+-\mu^-\)) spectra of the \cfo\ thin films with various thicknesses.
			Here, \(\mu^+\) (\(\mu^-\)) denote the absorption coefficient for photon helicity parallel (antiparallel) to the majority-spin direction.
			Figure \ref{fig:cfo_fe}(a) shows that the Fe \(L_{2,3}\)-edge XAS spectral line shapes \aoi{do not change significantly} by post-deposition annealing.
			In contrast, as shown in Fig. \ref{fig:cfo_fe}(b), the Fe \(L_{2,3}\)-edge XMCD intensities of \aoi{the} annealed samples dramatically increase compared to the as-grown ones, particularly in thinner films (\textit{d} = 1.4 and 2.3 nm).
			The increase of the XMCD \aoi{intensities} clearly demonstrates that \aoi{the post-deposition annealing recovers the ferrimagnetic order in the magnetically dead layer of the as-grown films.}
			The \textit{L}\textsubscript{3}-edge XMCD spectra show characteristic three peaks denoted by \textbf{a}, \textbf{b}, and \textbf{c}.
			It is well established that the three peaks \textbf{a}, \textbf{b}, and \textbf{c} originate from the Fe\(^{2+}(O_h)\), Fe\(^{3+}(T_d)\), and Fe\(^{3+}(O_h)\) ionic states, respectively \cite{Matzen2011, Moussy2013, Takaobushi2007, Matzen2014}.
			Therefore, one can discuss the Fe-ion distribution in spinel ferrites by estimating the intensities of these XMCD peaks.
			The spins of the Fe\(^{3+}(T_d)\) ions are antiferromagnetically coupled with those at the \(O_h\) sites, resulting in the opposite signs between peak \textbf{b} and peaks \textbf{a}/\textbf{c}.
			Due to these distinct peak features, XMCD spectra are more sensitive to the cation distribution in spinel ferrites than XAS \cite{Matzen2011,Moussy2013, Takaobushi2007}.
			Figures \ref{fig:cfo_fe}(c) and \ref{fig:cfo_fe}(d) show comparison of the Fe \textit{L}\textsubscript{3}-edge XMCD spectral line shapes of the as-grown and annealed \cfo\ films.
			The spectra have been normalized to the height of peak \textbf{c} to compare the spectral line shapes because the \(L_{2,3}\)-edge XMCD intensities decrease with the reduction of the film thickness due to the magnetically dead layers.
			Figures \ref{fig:cfo_fe}(b) and \ref{fig:cfo_fe}(c) show that the \(L_{2,3}\)-edge XMCD intensity decreases with the reduction of the thickness in the as-grown \cfo\ films and that the XMCD intensity at peak \textbf{b} decreases faster \aoi{than peaks \textbf{a} and \textbf{c}} as previously reported \cite{Wakabayashi2017}.
			This indicates the reduction of the Fe\(^{3+}(T_d)\) component of XMCD in magnetically dead layers near the interface.
			After post-deposition annealing, as shown in Fig. \ref{fig:cfo_fe}(b) and \ref{fig:cfo_fe}(d), both the entire XMCD intensities and the intensity at peak \textbf{b} are recovered simultaneously particularly in thinner films.
			This suggests that the Fe ions are redistributed and that ferrimagnetic order are simultaneously recovered \aoi{near the interface by annealing}.
			
			\begin{figure}
				\centering
				\includegraphics[width=0.5\linewidth]{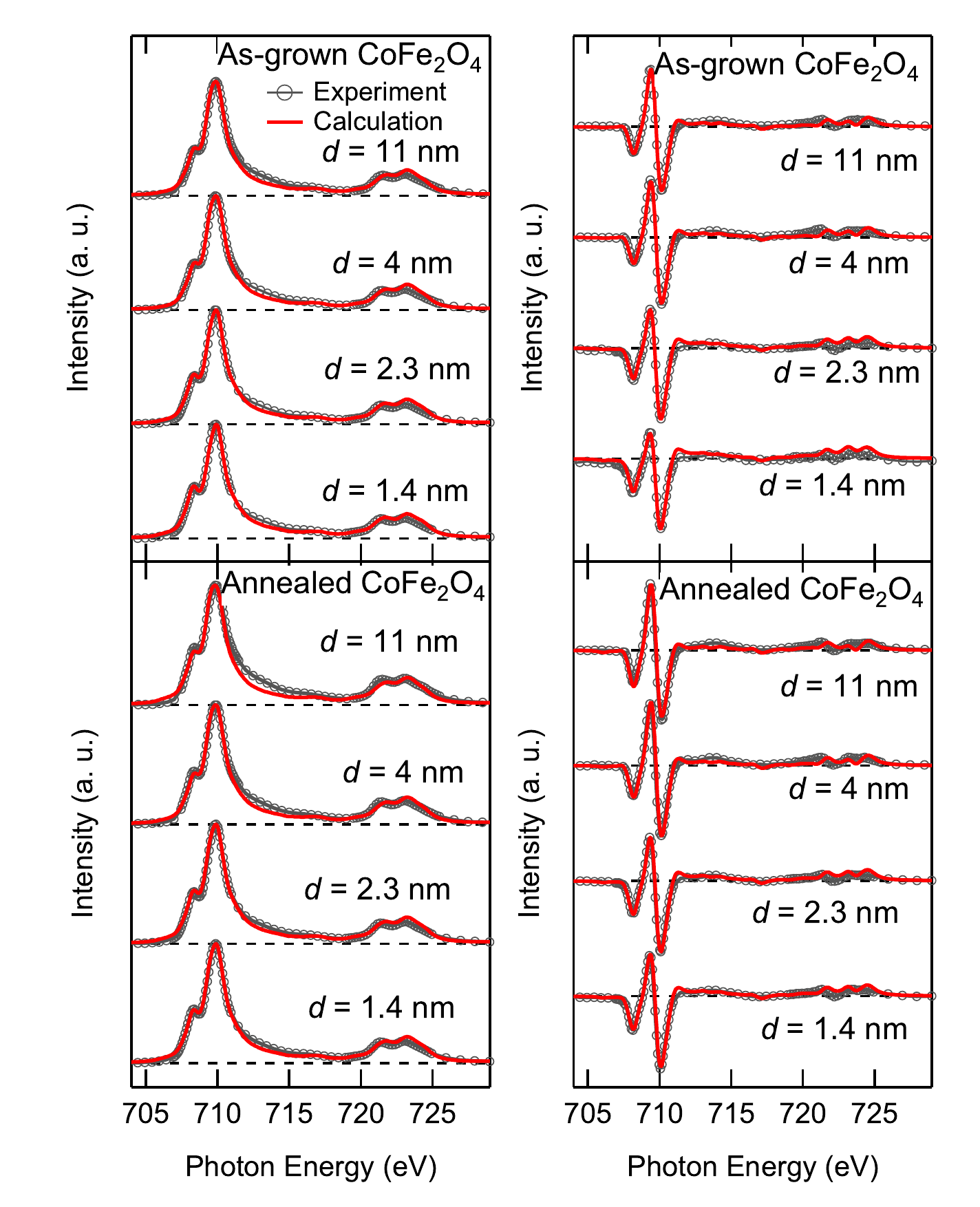}
				\caption{
					Comparison between experimental and calculated XAS and XMCD spectra at the Fe \(L_{2,3}\) edge of the as-grown and annealed \cfo\ thin films.
				}\label{fig:cfo_fits}
			\end{figure}
			
			\begin{figure}
				\centering
				\includegraphics[width=\linewidth]{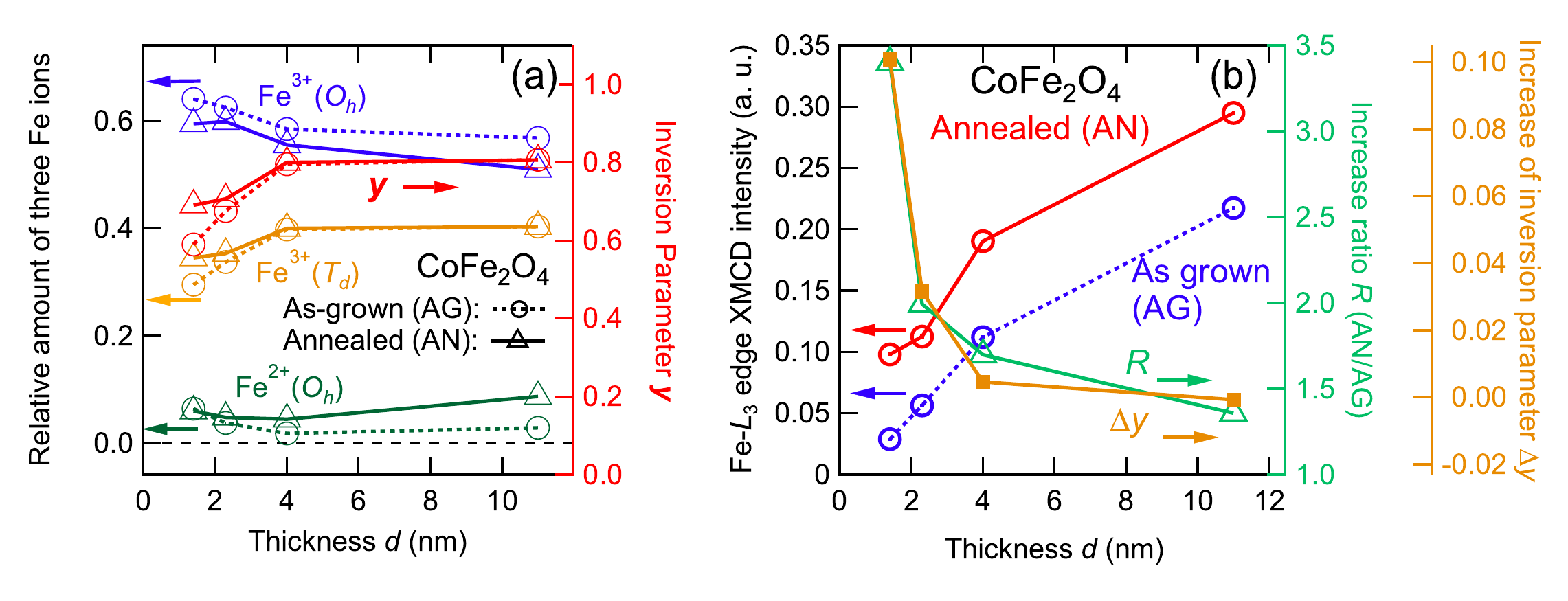}
				\caption{
					Thickness dependences of the Fe-ion distribution and the \(L_3\)-edge XMCD intensities for the as-grown and annealed \cfo\ thin films.
					(a) Relative amount of the Fe\(^{2+}(O_h)\), Fe\(^{3+}(T_d)\), and Fe\(^{3+}(O_h)\) ions as functions of film thickness. The inversion parameter \textit{y} is also plotted (right axis). Dashed lines and solid line indicate the as-grown and annealed samples, respectively.
					(b) Fe \(L_3\)-edge XMCD intensity of the as-grown and annealed films. The increase of \textit{y} by annealing (\(\Delta y\)) and the increase ratio of the XMCD intensity (\textit{R}) are plotted on the right axes.
					The inversion parameter \textit{y} decrease with reducing thickness in the as-grown samples [red dashed line in (a)], and \textit{y} increases by annealing [red solid line in (a)]. 
					\(\Delta y\) and \textit{R} similarly \aoi{decrease with} the film thickness [green and brown lines in (b)], suggesting that the recovery of the magnetically dead layer is achieved by the redistribution of cations.
				}\label{fig:cfo_fitsum}
			\end{figure}

			To clarify the correlation between the cation redistribution and the recovery of the ferrimagnetic order, quantitative spectral line-shape analyses are necessary.
			We, therefore, performed CI cluster-model calculation to reproduce the experimental XAS and XMCD spectra by the weighted sum of calculated Fe\(^{2+}(O_h)\), Fe\(^{3+}(T_d)\), and Fe\(^{3+}(O_h)\) spectra using the parameters listed in Table \ref{tab:para}.
			Figure \ref{fig:cfo_fits} shows comparison between the experimental and calculated XAS and XMCD spectra at the Fe \(L_{2,3}\) edge.
			All the experimental spectra are well reproduced by the calculated spectra.
			Since the amount of the ion is proportional to the weight for each calculated XAS and XMCD spectra, the relative amount can be calculated by the spectral weight ratio. 
			The relative amount deduced from fitting is plotted as functions of film thickness in Fig. \ref{fig:cfo_fitsum}(a), where the inversion parameter \textit{y} is also deduced and plotted.
			One can see that \textit{y} decreases with reducing thickness in the as-grown samples, and that \textit{y} increases by the post-deposition annealing. The increase of \textit{y} by annealing is more pronounced in the thinner samples\aka{, reflecting the redistribution of cations near the interface}.
			Figure \ref{fig:cfo_fitsum}(b) shows the Fe $L_3$-edge XMCD intensities of the as-grown and annealed films as functions of film thickness.
			The increase ratio of the Fe $L_3$-edge XMCD intensity [$R \equiv \text{XMCD (annealed)}/\text{XMCD (as grown)}$] and the increase of the inversion parameter [$\Delta y \equiv y \text{(annealed)} - y\text{(as grown)}$] are also shown for comparison.
			One can see similar thickness dependences between \textit{R} and \(\Delta y\), which suggests that the recovery of the magnetically dead layer is achieved by the redistribution of cations.
			Based on the above results and the definition of the inversion parameter \textit{y} of spinel ferrites (\([\text{Co}_{1-y}\text{Fe}_y]_{T_d}[\text{Fe}_{2-y}\text{Co}_y]_{O_h}\text{O}_4\)), one may think that the recovery of the magnetically dead layer by post-deposition annealing is caused by the reduction of the antisites near the interface, that is, the Co ions irregularly occupying the \(T_d\) sites are replaced by Fe ions after annealing. However, one also needs to take into account the cation vacancies at the \(T_d\) sites, as discussed below. 
			
			\begin{figure}
				\centering
				\includegraphics[width=0.5\linewidth]{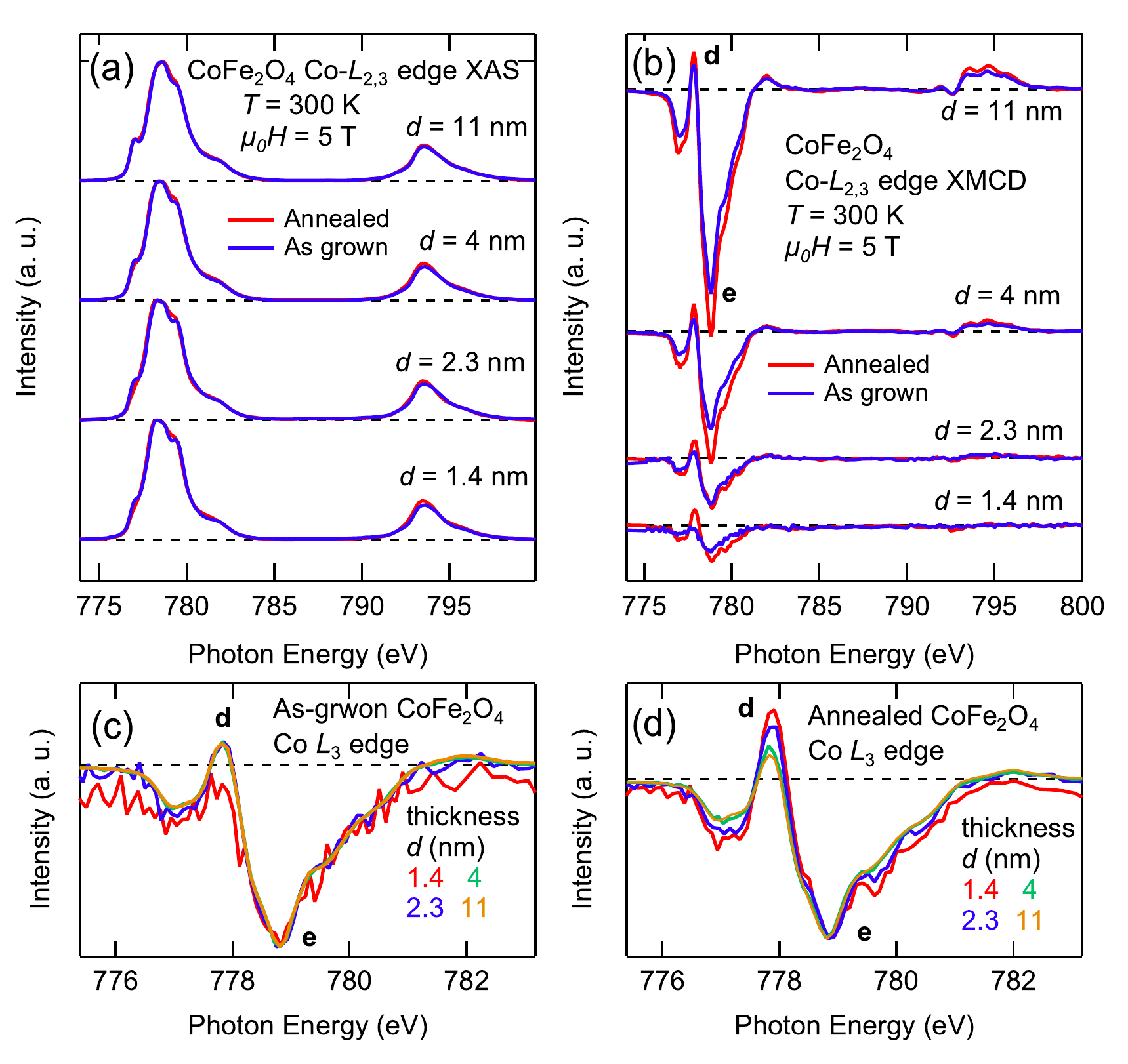}
				\caption{
					Co \textit{L}\textsubscript{2,3}-edge XAS and XMCD spectra of the as-grown and annealed \cfo\ thin films with various thicknesses. 
					(a), (b) Co \textit{L}\textsubscript{2,3}-edge XAS and XMCD spectra. 
					(c), (d) Magnified views of Co \textit{L}\textsubscript{3}-edge XMCD spectra normalized to the height of peak \textbf{e}.
					Since peak \textbf{d} originates from Co\(^{2+}(T_d)\) ions \cite{Wakabayashi2017}, the increase of peak \textbf{d} indicates that the number of ferrimagnetically ordered Co\(^{2+}(T_d)\) ions increase by post-deposition annealing.
				}\label{fig:cfo_co}
			\end{figure}
			Figure \ref{fig:cfo_co} shows the Co \(L_{2,3}\)-edge XAS and XMCD spectra for various film thicknesses and the effect of annealing on these spectra.
			In order to see the changes in the spectral line shapes, we show the Co-\(L_3\) edge XMCD spectra normalized to the height of peak \textbf{e} in Figs. \ref{fig:cfo_co}(c) and \ref{fig:cfo_co}(d).
			One can see that the positive peak \textbf{d} increases by annealing, particularly in the thinner films.
			As described in the previous report \cite{Wakabayashi2017}, the positive peak at the Co \(L_3\) edge originates from Co\(^{2+}(T_d)\) ions. 
			Therefore, the increase of peak \textbf{d} indicates the increase of the number of ferrimagnetically ordered Co\(^{2+}(T_d)\) ions near the interface after annealing.
			The increase of the positive peak components could also explain the slight increase of the Co \(L_{2,3}\)-edge XMCD intensities by annealing. 
			Surprisingly, not only the number of Fe\(^{3+}(T_d)\) ions but also that of ferrimagnetically ordered Co\(^{2+}(T_d)\) ions increase by annealing.
			These results can be naturally understood if there are a significant number of vacant \(T_d\) sites near the interface of the as-grown \cfo\ thin films and that, after annealing, the vacancies are filled by Fe ions and possibly Co ions, too, diffused from the \(O_h\) sites (for details, see Section \ref{anneal_discussion}).
			We note that the inversion parameter \textit{y} is defined based on the ideal spinel structure (\([\text{Co}_{1-y}\text{Fe}_y]_{T_d}[\text{Fe}_{2-y}\text{Co}_y]_{O_h}\text{O}_4\)) and the vacant sites are not taken into account.
			The lowering of \textit{y} may reflect not only the replacement of the Fe ion by the Co ion but also increase of the vacancies at the \(T_d\) and/or \(O_h\) sites.

		\subsection{\nfo\ thin films}\label{anneal_nfo}
			\begin{figure} 
				\centering
				\includegraphics[width=0.5\linewidth]{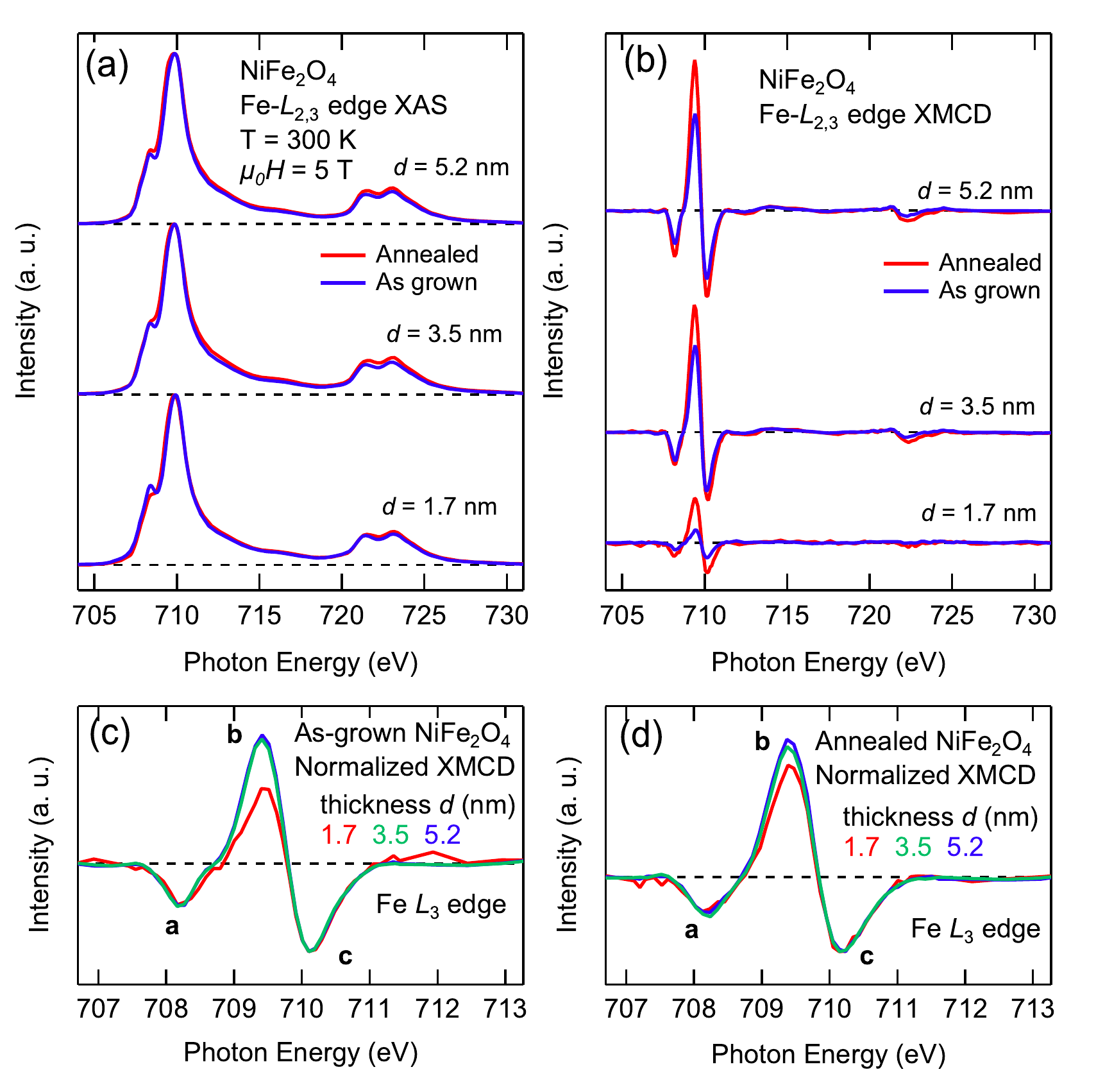}
				\caption{
					Fe \textit{L}\textsubscript{2,3}-edge XAS and XMCD spectra of the as-grown and annealed \nfo\ thin films with various thicknesses \(d=\text{1.7, 3.5, and 5.2 nm}\). 
					(a), (b) Fe \(L_{2,3}\)-edge XAS and XMCD spectra. 
					(c), (d) Magnified views of the Fe \(L_3\)-edge XMCD spectra normalized to the height of peak \textbf{c}.
					The changes in the spectra are similar to those of \cfo\ (Fig. \ref{fig:cfo_fe}).
				}\label{fig:nfo_fe}
			\end{figure}
\begin{figure} 
				\centering
				\includegraphics[width=0.5\linewidth]{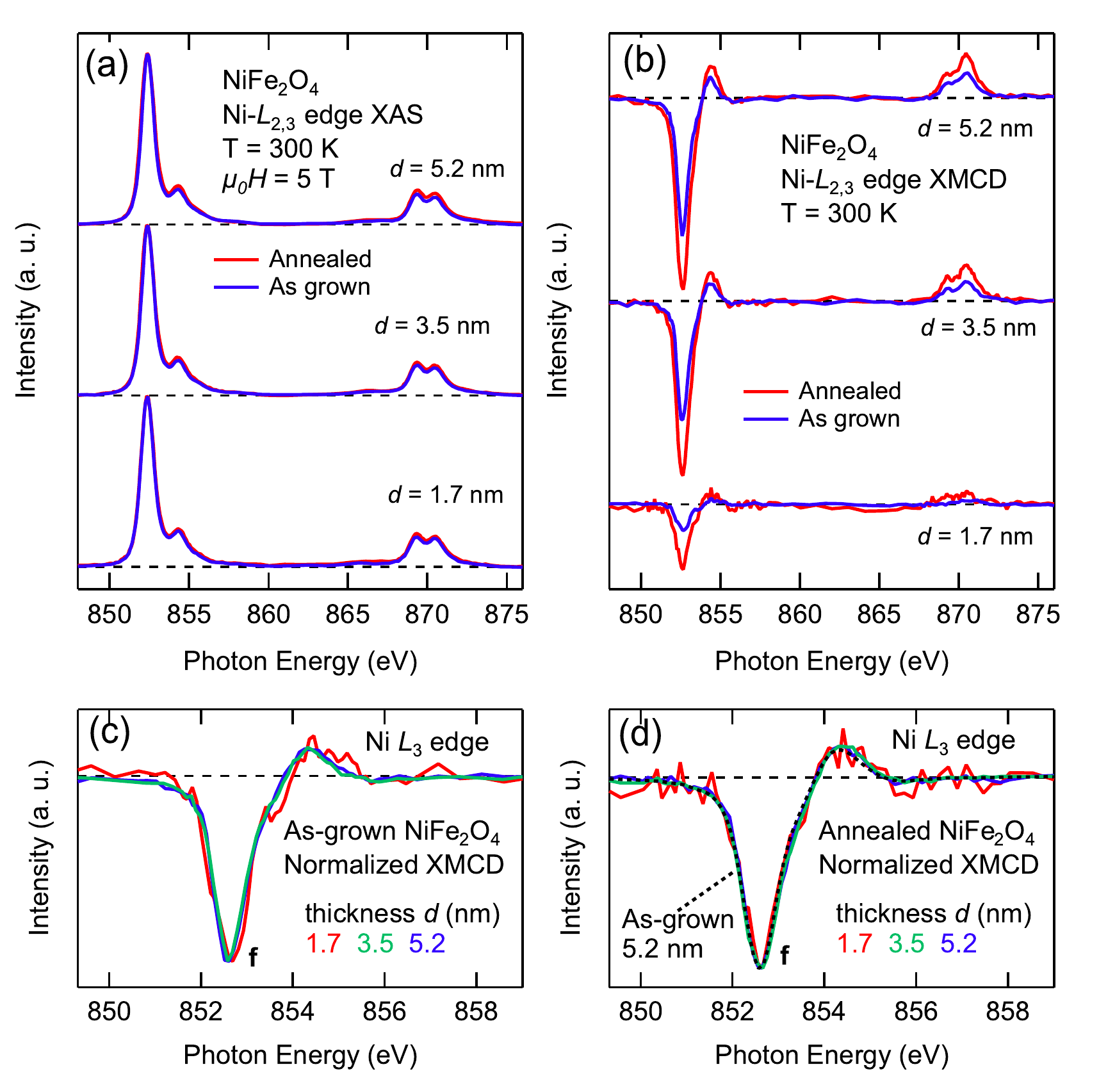}
				\caption{
					Ni \textit{L}\textsubscript{2,3}-edge XAS and XMCD spectra of the as-grown and annealed \nfo\ thin films with various thicknesses \(d=\text{1.7, 3.5, and 5.2 nm}\). The spectra of the as-grown films (blue lines) are adopted from Ref. \onlinecite{Wakabayashi2018}.
					(a), (b) Ni \(L_{2,3}\)-edge XAS and XMCD spectra. 
					(c), (d) Magnified views of the Ni \(L_3\)-edge XMCD spectra normalized to the height of peak \textbf{f} for the as-grown and annealed films. The normalized XMCD spectra of the 5.2-nm-thick as-grown film is also shown by a dashed black curve for comparison. 
					In contrast to the Co \(L_{2,3}\)-edge XMCD spectra shown in Fig. \ref{fig:cfo_co}, spectral line shapes are not changed by annealing, reflecting the \aka{high} \aoi{\(O_h\)-}site selectivity of the Ni\(^{2+}\) ion.
				}\label{fig:nfo_ni}
			\end{figure}

			Figure \ref{fig:nfo_fe} shows the Fe \(L_{2,3}\)-edge XAS and XMCD spectra of the as-grown and annealed \nfo\ thin films with thicknesses \(d=1.7, 3.5,\text{ and } 5.2\) nm.
			Figure \ref{fig:nfo_ni} shows the Ni \(L_{2,3}\)-edge XAS and XMCD spectra of the \nfo\ films.
			The XMCD intensity of the 1.7 nm-thick film is extremely low compared to the others, indicating the dominant effect of magnetically dead layers.
			After annealing, the XMCD intensities increase for all the films.
			This clearly indicates that the magnetic order in the dead layers was significantly recovered by annealing also in the \nfo\ thin films.
			%

			To see the effect of annealing on the cation distribution, we compare the line shapes of normalized XMCD spectra at the Fe and Ni \(L_3\) edges. 
			Figures \ref{fig:nfo_fe}(c) and \ref{fig:nfo_fe}(d) show magnified views of the Fe \(L_3\)-edges XMCD spectra normalized to the height of peak \textbf{c}.
			The spectral line shapes show that the positive peak \textbf{b} of the 1.7 nm-thick film significantly increases after annealing, indicating the increase of Fe\(^{3+}(T_d)\) ions.
			On the other hand, the Ni-edge XMCD spectral line shapes, shown in Figs. \ref{fig:nfo_ni}(c) and \ref{fig:nfo_ni}(d), do not show any visible changes depending on the thickness and annealing.
			The XMCD spectra of the 5.2 nm-thick as-grown film are shown in Fig. \ref{fig:nfo_ni}(d) as a dashed black curve for comparison.
			One can clearly see that the spectral line shape remains the same after annealing.
			Because the Ni ion takes only the Ni\(^{2+}(O_h)\) state, as reported in as-grown films \cite{Wakabayashi2018}, the unchanged spectral line shapes by annealing indicate that the Ni ions did not move to the \(T_d\) sites even after annealing, confirming the high \aoi{\(O_h\)-}site selectivity of the Ni\(^{2+}\) ion.
			
			\begin{figure} 
				\centering
				\includegraphics[width=0.5\linewidth]{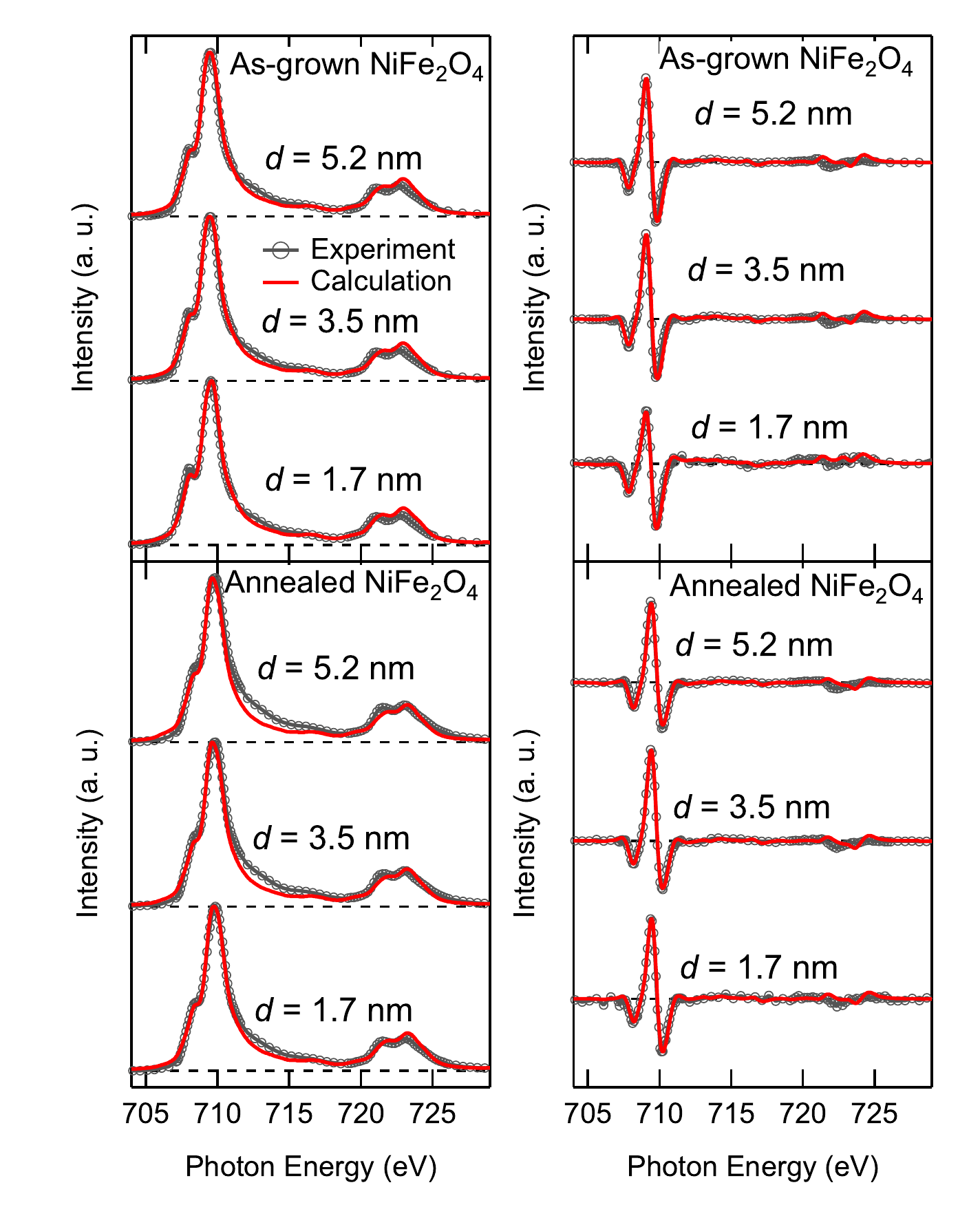}
				\caption{
					Comparison between the experimental and calculated Fe \(L_{2,3}\)-edge XAS and XMCD spectra of the as-grown and annealed \nfo\ thin films.
				}\label{fig:nfo_fits}
			\end{figure}
\begin{figure} 
				\centering
				\includegraphics[width=\linewidth]{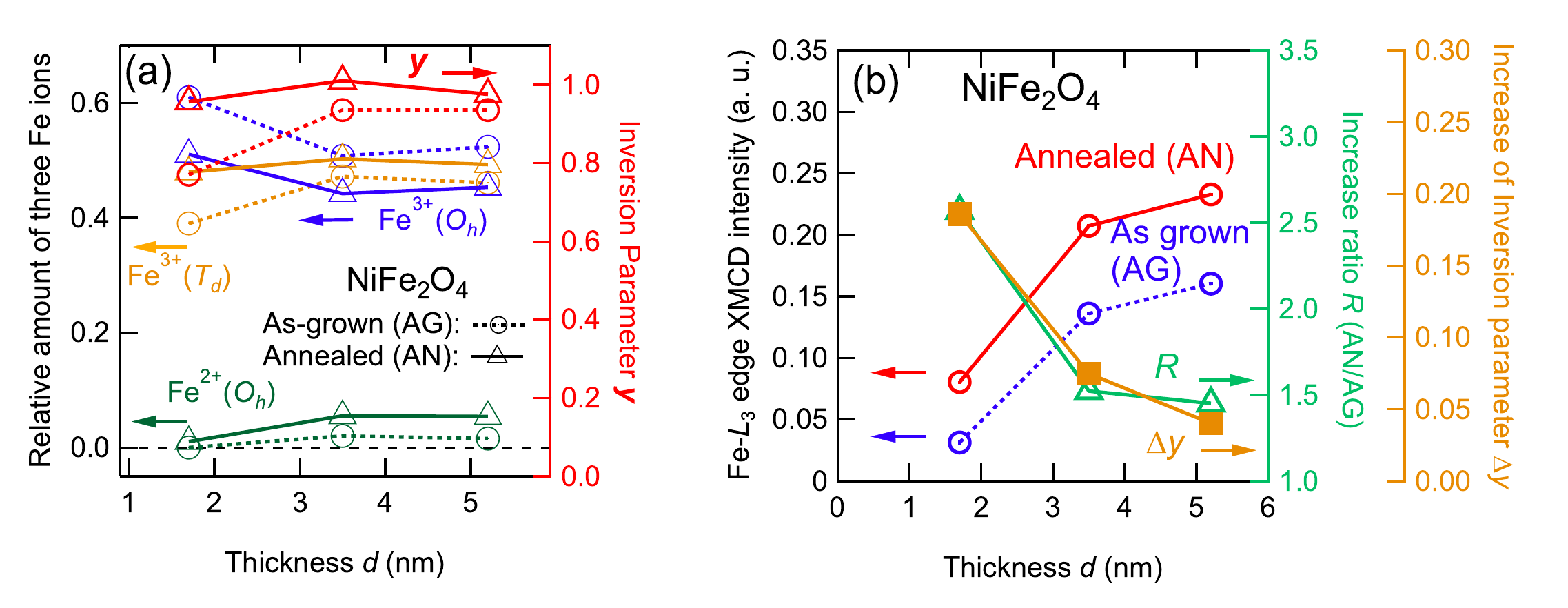}
				\caption{
					Thickness dependences of the Fe-ion distribution and the \(L_3\)-edge XMCD intensities for the as-grown and annealed \nfo\ thin films.
					(a) Relative amounts of the Fe\(^{2+}(O_h)\), Fe\(^{3+}(T_d)\), and Fe\(^{3+}(O_h)\) ions in the as-grown and annealed \nfo\ thin films as functions of thickness. The inversion parameter \textit{y} is also plotted on the right axis. Dashed lines with open circles and solid lines with open triangles indicate as-grown and annealed samples, respectively.
					(b) Fe \(L_3\)-edge XMCD intensity of the as-grown and annealed films. The increase of \textit{y} by annealing (\(\Delta y\)) and the increase ratio of XMCD intensities (\textit{R}) are plotted on the right axes.
				}\label{fig:nfo_fitsum}
			\end{figure}

			In order to quantitatively analyze the spectral line shapes of \nfo, as in the case of \cfo, we fitted the measured Fe \(L_{2,3}\)-edge XAS and XMCD spectra using the weighted sum of the three calculated spectra of Fe\(^{2+}(O_h)\), Fe\(^{3+}(T_d)\), and Fe\(^{3+}(O_h)\) ions.
			The parameters for the CI cluster model were the same as those for \cfo\ listed in Table \ref{tab:para}.
			Figure \ref{fig:nfo_fits} shows the experimental and calculated Fe \(L_{2,3}\)-edge XAS and XMCD spectra.
			The experimental spectra are well reproduced by the calculation.
			Using the weights obtained from the fitting, one can estimate the relative amount of the Fe\(^{2+}(O_h)\), Fe\(^{3+}(T_d)\), and Fe\(^{3+}(O_h)\) ions in the as-grown and annealed \nfo\ thin films as shown in Fig. \ref{fig:nfo_fitsum}(a).
			One can see that, for the 1.7-nm-thick film, the ratio of the Fe\(^{3+}(T_d)\) ions increases with post-deposition annealing, while that of the Fe$^{3+}$($O_h$) ions decreases.
			Thus, the inversion parameter \textit{y} of the as-grown 1.7 nm-thick film is low compared to those of the thicker films, consistent with the XMCD result.
			The low \textit{y} value of the 1.7 nm-thick film significantly increases to 0.96 after annealing (red dashed line).
			This means that NiFe$_2$O$_4$ thin films with almost no antisite defects are achieved after annealing.
			Figure \ref{fig:nfo_fitsum}(b) shows the Fe $L_3$-edge XMCD intensities of the as-grown and annealed films as functions of film thickness. The increasing ratio of the Fe $L_3$-edge XMCD intensity \textit{R} and the increase of the inversion parameter \(\Delta y\) are also shown for comparison.
			Analogously to \cfo\ shown in Fig. \ref{fig:cfo_fitsum}(b), the similar thickness dependences between $R$ and $\Delta y$ suggest that the recovery of the magnetically dead layer is achieved by the redistribution of cations.
			
			The ratio of Fe\(^{3+}(T_d)\) to Fe\(^{3+}(O_h)\) was about \(\frac{2}{3}\) in the 1.7 nm-thick film before annealing. 
			Considering that the Ni ions are located only at the \(O_h\) sites, one can estimate from this ratio that about \(\frac{1}{3}\) of the \(T_d\) sites should be vacant in the as-grown 1.7 nm-thick film.
			As we shall discuss below, these vacancies at the \(T_d\) sites should have detrimental effects on the ferrimagnetic order in the spinel structure.

	\section{Discussion}\label{anneal_discussion}
		We discuss the role of the cation vacant sites in the magnetically dead layers.
		As introduced above, \aoi{based on the studies on Fe\textsubscript{3}O\textsubscript{4} \cite{Margulies1996, Margulies1997, Moussy2004}}, it is \aoi{believed} that the magnetically dead layers in the spinel ferrites are caused by antiphase boundaries.
		However, as shown above, we find that there are a significant number of vacant \(T_d\) sites near the interface, and the transfer of the cations from the \(O_h\) to \(T_d\) sites caused by annealing significantly recovers the ferrimagnetic order.
		The antiferromagnetic superexchange interaction between the \(T_d\)-site and \(O_h\)-site cations is the origin of the ferrimagnetic order in the spinel ferrites.
		Therefore, the lack of superexchange interaction due to the vacancies should significantly suppress the ferrimagnetism of magnetically dead layers.
		Here, we focus on how the vacant \(T_d\) sites affect the spin structures in the magnetically dead layers.

		\begin{figure} 
			\centering
			\includegraphics[width=0.8\linewidth]{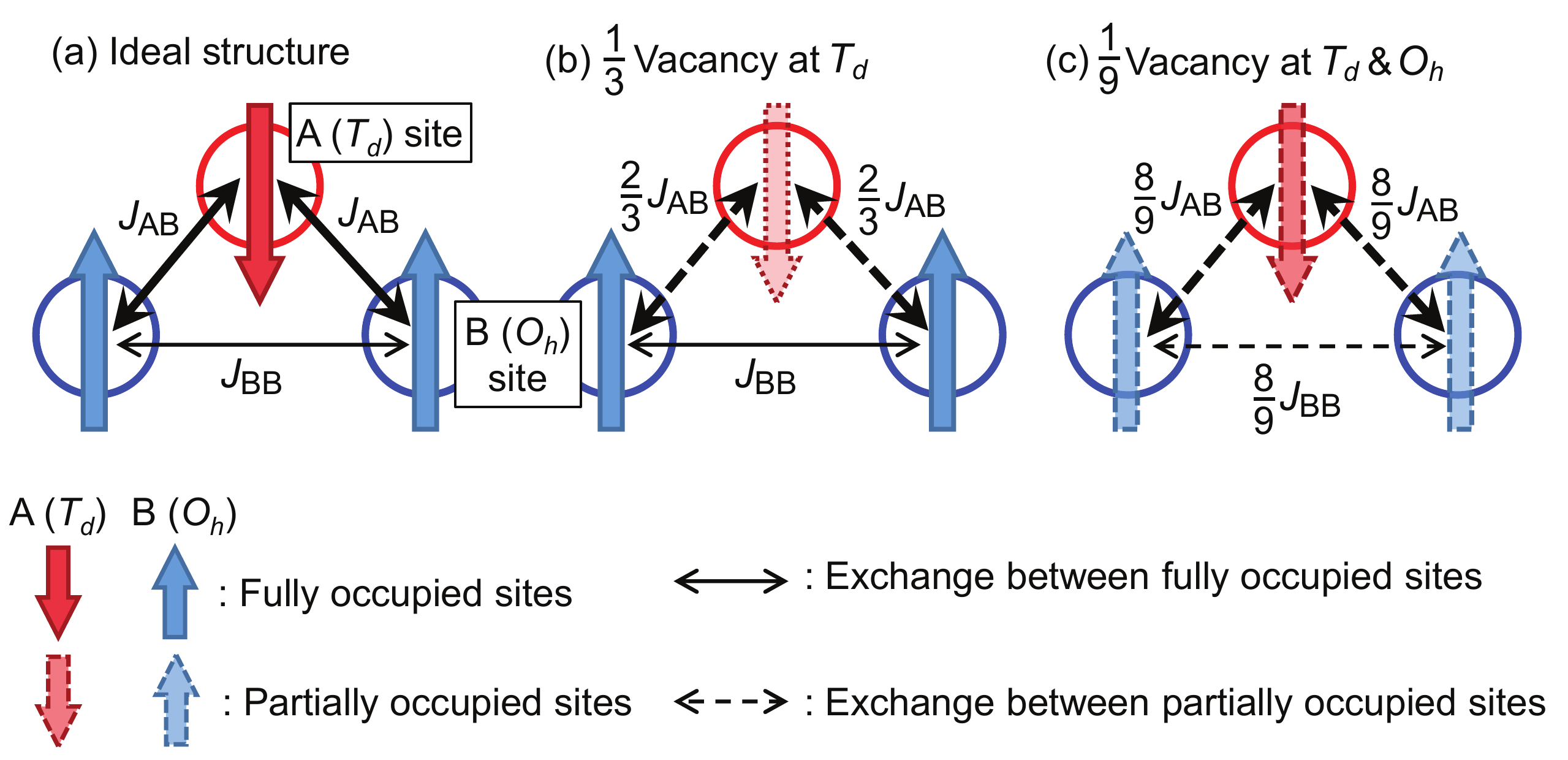}
			\caption{
				Schematic pictures of exchange interaction between the \(T_d\)- and \(O_h\)-site ions.
				The \(T_d\) and \(O_h\) sites are labeled \(A\) and \(B\), respectively. 
			}\label{fig:discussion}
		\end{figure}
		\begin{table} 
			\caption{
			Exchange interaction between transition metal ions in \(\textit{M}\text{Fe}_2\text{O}_4\) (\textit{M}=\text{Fe, Co}) in units of K reproduced from Ref. \onlinecite{Srivastava1979}.
			Suffixes \(A\) and \(B\) denotes the \(T_d\) and \(O_h\) sites, respectively.
			Note that the listed values include both the superexchange and direct exchange interactions.
			Positive and negative values correspond to ferromagnetic and antiferromagnetic interactions, respectively.
			}\begin{center}
			\scalebox{1.0}{ 
				\begin{tabular}{c c | c c} 
				\hline
				\hline
					&	&\textit{M} = Co	&\textit{M} = Ni	\\
				\hline
				\multirow{2}{*}{\(J_{{AB}}\)}	&\(\text{Fe}^{3+}\)-\(\textit{M}^{2+}\)	&\(-22.7\)	&\(-27.4\)	\\
					&\(\text{Fe}^{3+}\)-\(\text{Fe}^{3+}\)	&\(-26\)	&\(-30.7\)	\\
				\hline
				\multirow{3}{*}{\(J_{{BB}}\)}	&\(\textit{M}^{2+}\)-\(\textit{M}^{2+}\)	&\(46.9\)	&\(30.0\)	\\
					&\(\text{Fe}^{3+}\)-\(\textit{M}^{2+}\)	&\(-18.5\)	&\(-2.7\)	\\
					&\(\text{Fe}^{3+}\)-\(\text{Fe}^{3+}\)	&\(-7.5\)	&\(-5.4\)	\\
				\hline
				\hline
				\end{tabular}
			}\label{tab:exchange}
			\end{center}
		\end{table}
		Figure \ref{fig:discussion}(a) shows the exchange interaction between the cations at the \(O_h\) and \(T_d\) sites of the spinel structure.
		The values of the exchange interaction in \(\textit{M}\text{Fe}_2\text{O}_4\) (\textit{M}=\text{Co, Ni}) deduced from magnetization measurements \cite{Srivastava1979}, which are denoted as \(J_{AB}\) and \(J_{BB}\) in Fig. \ref{fig:discussion}, are listed in Table \ref{tab:exchange} in units of K.
		We note that suffix \(A\) (\(B\)) denotes the \(T_d\) (\(O_h\)) sites, and the ratio between \(M^{2+}(O_h)\), Fe\(^{3+}(T_d)\), and Fe\(^{3+}(O_h)\) is \(1:1:1\) in an ideal inverse spinel structure.
		The large negative values of \(J_{AB}\) demonstrate the strong contributions of the \(T_d\) site to the ferrimagnetic order.
		As for the \(O_h\)-\(O_h\) interaction, \(J_{BB}\), one can see large positive values for \(M^{2+}\)-\(M^{2+}\) interaction, and negative ones for \(\text{Fe}^{3+}\)-\(\text{Fe}^{3+}\) and \(\text{Fe}^{3+}\)-\(M^{2+}\) interactions.
		On average, \(O_h\)-\(O_h\) interactions are weakly ferromagnetic (\(+0.6\) K for \cfo\ and \(+4.8\) K for \nfo) \footnote{
			The number of \(\text{Fe}^{3+}\)-\(M^{2+}\) exchange pairs is twice as much as that of \(M^{2+}\)-\(M^{2+}\) and \(\text{Fe}^{3+}\)-\(\text{Fe}^{3+}\) exchange pairs.
			Therefore, the weighted average value is \(\frac{1}{4} (J_{BB}^{M^{2+}-M^{2+}}+2J_{BB}^{\text{Fe}^{3+}-M^{2+}}+J_{BB}^{\text{Fe}^{3+}-\text{Fe}^{3+}})\).}.
		We note that the number of the \(O_h\)-\(T_d\) exchange pairs in an ideal spinel structure are 2.3 times larger than that of \aoi{the} \(O_h\)-\(O_h\) ones \cite{Margulies1997}.
		If \(\frac{1}{3}\) of the \(T_d\) site is vacant as shown in Fig. \ref{fig:discussion}(b), \(J_{AB}\) is weaken to \(\frac{2}{3}\) of the ideal value, degrading the ferrimagnetic order.
		After annealing, some \(O_h\)-site ions move to \(T_d\) sites. 
		For simplicity, assuming that the vacancies at the \(T_d\) sites are equally redistributed to all the \(T_d\) and \(O_h\) sites, as shown in Fig. \ref{fig:discussion}(c), antiferromagnetic \(J_{AB}\) is recovered to \(\frac{8}{9}\) of ideal value.
		Since the value and number of exchange interactions of \(J_{AB}\) are larger than that of \(J_{BB}\), ferrimagnetic order should be stabilized.
		Although the actual numbers of the vacancies are still unclear, it can be concluded that vacancies at the \(O_h\) sites still preserve the ferrimagnetic order, whereas vacancies at the \(T_d\) sites significantly degrade the ferrimagnetic order.
		We should note that the above discussion ignores the effects of charge unbalance caused by the cation vacancies, that is, the effects of cation valence changes and possible oxygen vacancies.
		We speculate that there are oxygen vacancies \cite{Jaffari2012} and/or that extra \(O_h\) sites \cite{Chang2016} near the interface.
			
	\section{Conclusion}
		We have studied the electronic and magnetic states of the as-grown and annealed \cfo(111) and \nfo(111) thin films with various thicknesses grown on Si substrates with the \(\gamma\)-\alo\ buffer layers.
		By analyzing the spectral line shapes using the CI cluster-model calculation, we have estimated the valences and site occupancies of Fe ions.
		We confirmed through the significant reduction of the XMCD intensities \aoi{with decreasing film thickness} that there are magnetically dead layers near the interface with $\gamma$-Al$_2$O$_3$.
		As reported in the previous works \cite{Wakabayashi2017, Wakabayashi2018}, the reduction of the inversion parameters was observed in the magnetically dead layers.
		By annealing the films for 48 hours in air to redistribute the cations, the XMCD intensity and inversion parameters significantly increase particularly in the thinner films.
		Furthermore, from comparison between the results on the as-grown and annealed films, it is suggested that there are a significant number of vacant \(T_d\) sites in the magnetically dead layers.
		After the post-deposition annealing, the vacancies at the \(T_d\) sites are partially occupied by ions \aoi{diffused} from the \(O_h\) sites.
		Thus, we conclude that vacant \(T_d\) sites play an important role in the formation of the magnetically dead layers in spinel ferrites.
	\section*{Acknowledgments}
		We would like to thank Ryosho Nakane for valuable advice and enlightening discussion throughout the present work, Kenta Amemiya and Masako Suzuki-Sakamaki for valuable technical support at KEK-PF BL-16A.
		This work was supported by KAKENHI grants from JSPS (JP15H02109, JP26289086, JP15K17696, JP19K03741, JP20K14416, JP20H02199, and JP22K03535).
		The experiment was done under the approval of the Photon Factory Program Advisory Committee (Proposal No. 2016S2-005).
		Y. K. W. and Z. C. acknowledges financial support from Materials Education Program for the Futures leaders in Research, Industry and Technology (MERIT). 
		Y. W. acknowledges financial support from Advanced Leading Graduate Course for Photon Science (ALPS). 
		Y. K. W. also acknowledge support from the JSPS Research Fellowship Program for Young Scientists.

	\bibliography{../library.bib}
\end{document}